\documentclass[%
 aip,
 jmp,%
 amsmath,amssymb,
 reprint,%
]{revtex4-1}

\usepackage{graphicx}
\usepackage{dcolumn}
\usepackage{bm}

\begin{document}

\preprint{AIP/123-QED}

\title{External electric field effect on electron transport in carbon nanotubes}

\thanks{Corresponding author: sulemana70@gmail.com
}

\author{S. S. Abukari}%
\affiliation{Department of Physics, Laser and Fibre Optics Centre, University of Cape Coast, Cape Coast, Ghana.
}%

\author{S. Y. Mensah}
\affiliation{Department of Physics, Laser and Fibre Optics Centre, University of Cape Coast, Cape Coast, Ghana.
}%

\author{N. G. Mensah}
\affiliation{Department of Mathematics, University of Cape Coast, Cape Coast, Ghana.
}%

\author{M. Rabiu}
 \affiliation{Department of Applied Physics, Faculty of Applied Sciences, University for Development Studies, Navrongo Campus, Ghana.}

\author{K. W. Adu}
\affiliation{Department of Physics, The Pennsylvania State University Altoona College, Altoona, Pennsylvania 16601, USA. Research Institute, The Pennsylvania State University, University Park, PA 16802, USA
}%

\author{A. Twum}
\affiliation{Department of Physics, Laser and Fibre Optics Centre, University of Cape Coast, Cape Coast, Ghana.
}%

\date{\today}

\begin{abstract}
Electronic transport properties of carbon nanotubes are studied theoretically in the presence of external electric field $E(t)$ by using the Boltzmann's transport with constant relaxation time. An analytical expression for the current densities of the nanotubes are obtained. It is observed that the current density-electric field characteristics of the CNs exhibit total self-induced transparency and absolute negative conductivity
%
\end{abstract}



\maketitle

\section{Introduction}
The electrical transport properties of carbon nanotubes (CNs) have been the subject of much research ever since the discovery by Iijima \cite{1} of the quasi-one-dimensional monomolecular structures. This may be due to their abilities to exhibit Bloch oscillations \cite{2,5} at moderate electric field strengths. This oscillatory response makes CNs inherently nonlinear and as such can perform varieties of transport phenomena. Under different conditions of an external electric field, an electron is predicted to reveal a variety of physical effects such as Bloch oscillations, self-induced transparency, negative differential conductivity, absolute negative conductance \cite{4}, etc.

We shall demonstrate in this paper two phenomena in CNs, which are self-induced transparency and absolute negative conductance for the following cases respectively:
\begin{itemize}
	\item When the CNs is exposed to an a.c electric field, i.e $E(t) = E_1 cos\omega t$
	\item When the CNs is exposed to an a.c. and d.c electric field, i.e $E(t) = E_0 + E_1 cos\omega t$
\end{itemize}

\section{Theory}
Following ref. \cite{3, 4} and using the approach similar to ref. \cite{6}, we consider a response of electrons in an undoped achiral single-wall carbon nanotubes subject to an external electric field,  
\begin{equation}
	E(t)=E_0 + E_1 cos\omega t \label{eq:one}
\end{equation}

We use the semiclassical approximation in which $\pi$-electrons are considered as classical particles with dispersion law extracted from the quantum theory in the tight-binding approximation \cite{4}. 

Considering the hexagonal crystalline structure of zigzag and armchair CNTs within tight binding approximation, the dispersion relation is given as \cite{4} respectively,
\begin{eqnarray}
	\epsilon_s(p_z) &=& \pm\gamma_0\Big[ 1 + 4 cos (ap_z) cos\Big(\frac{a}{\sqrt{3}}s\Delta p_\varphi\Big) \ldots\nonumber \\
	 && + 4cos^2\Big( \frac{a}{\sqrt{3}}s\Delta p_\varphi\Big)\Big]^{1/2}\label{eq:two}
\end{eqnarray}   
\begin{eqnarray}
	\epsilon_s(p_z) &=& \pm\gamma_0\Big[ 1 + 4 cos (as\Delta p_{\varphi}) cos\Big(\frac{a}{\sqrt{3}} p_z\Big) \ldots\nonumber\\
	 && + 4 cos^2\Big(\frac{a}{\sqrt{3}} p_z\Big)\Big]^{1/2}\label{eq:three}
\end{eqnarray}
Here $\gamma_0 \sim 3.0$ eV is the overlapping integral, $p_z$ is the axial component of quasimomentum, $\Delta p_{\varphi}$ is transverse quasimomentum level spacing and $s$ is an integer. The expression for  $a$ in Eqns. \eqref{eq:two} and  \eqref{eq:three} is given as $a=3b/2\hbar$. With the C-C bond length  $b=0.142$ nm  and  $\hbar$ is the Plank's constant, we shall assume $\hbar=1$. The $-$ and $+$ signs correspond to the valence and conduction bands, respectively. Due to the transverse quantization of the quasi-momentum, its transverse component can take $n$ discrete values, $p_{\varphi} = s\Delta p_{\varphi} = (\pi\sqrt{3}s)/an$, ($s=1, \ldots, n$). Unlike transverse quasimomentum $p_{\varphi}$, the axial quasimomentum $p_z$ is assumed to vary continuously within the range $0 \leq p_z \leq 2\pi/a$, which corresponds to the model of infinitely long CNT ($L = \infty$). This model is applicable to the case under consideration because of the restriction to the temperatures and /or voltages well above the level spacing \cite{4}, i.e, $k_B T > \epsilon_C$, $\Delta\epsilon$, where $k_B$ is Boltzmann constant, $T$ is the temperature, $\epsilon_C$ is the charging energy. The energy level spacing $\Delta\epsilon$ is given by $\Delta\epsilon = \pi\hbar v_F/L$, where $v_F$ is the Fermi velocity and $L$ is the carbon nanotube length \cite{6}

Employing Boltzmann equation with relaxation time $\tau = constant$, 
\begin{equation}
	\frac{\partial f(p,t)}{\partial t} + eE(t)  \frac{\partial f(p,t)}{\partial t} =-\frac{f_s-f_0 (p)}{\tau} \label{eq:four}
\end{equation}
where $e$ is the electron charge, $f_0 (p)$ is the equilibrium distribution function, $f(p,t)$ is the distribution function, and $\tau$ is the relaxation time. The electric field $E(t)$ is applied along CNTs axis. In this problem the relaxation term $\tau$ is assumed to be constant.
 
Expanding the distribution functions of interest in Fourier series as;
\begin{equation}
	f_0(p) = \Delta p_{\varphi} \sum_{s=1}^{n}\delta(p_{\varphi} - s\Delta p_{\varphi})\sum_{r\neq 0} f_{rs}e^{iarp_z} \label{eq:five}
\end{equation}
and
\begin{equation}
	f(p,t) = \Delta p_{\varphi} \sum_{s=1}^{n}\delta(p_{\varphi} - s\Delta p_{\varphi})\sum_{r\neq 0} f_{rs}e^{iarp_z}\Phi_{\nu} \label{eq:six}
\end{equation}
Here the coefficient, $\delta(x)$ is the Dirac delta function, $f_{rs}$ is the coefficient of the Fourier series and $\Phi_{\nu}(t)$ is the factor by which the Fourier transform of the nonequilibrium distribution function differs from its equilibrium distribution counterpart. The expression $f_{rs}$ can be expanded in the analogous series as follows
\begin{equation}
 	  f_{rs} = \frac{a}{2\pi}\int_0^{\frac{2\pi}{a}} \frac{e^{-iarp_z}}{1 + exp(\epsilon_s(p_z))/k_BT)}dp_z \label{eq:seven}
\end{equation}
Substituting  Eqns. \eqref{eq:six} and \eqref{eq:seven} into Eqn. \eqref{eq:three}, and solving with Eqn. \eqref{eq:one} we obtain
\begin{equation}
	\Phi_{\nu}(t) = \sum_{k=-\infty}^{\infty}\sum_{\nu=-\infty}^{\infty}\frac{J_k(r\beta)J_{k-\nu}(r\beta)}{1 + i(earE_0 + k\omega)\tau}exp(i\nu\omega t). \label{eq:eight}
\end{equation}
where  $\beta = (eaE_1)/\omega$, $J_k (r\beta)$ is the Bessel function of the $k^{th}$ order and $\Omega = eaE_0$. 

Similarly, taking into account the relation $v_z (p_z,s\Delta p_{\varphi} ) = \partial\epsilon_{rs} (p_z )/\partial p_z $, we represent $\epsilon_s (p_z)/\gamma_0$ in Fourier series with the coefficients as follows;
\begin{equation}
	\epsilon_{rs} = \frac{a}{2\pi\gamma_0}\int_0^{\frac{2\pi}{a}}\epsilon_s(p_z)e^{-irap_z}dp_z, \label{eq:nine}
\end{equation}
the quasiclassical velocity $v_z (p_z,s\Delta p_z )$ of an electron moving along the CNs axis is given by the expression
\begin{equation}
	v_z (p_z, s?p_z ) = \gamma_0 \sum_{r \neq 0} \frac{\partial (\epsilon_{rs} e^{iarp_z} )}{\partial p_z } = \gamma_0 \sum_{r  \neq 0} iar \epsilon_{rs} e^{iarp_z}     \label{eq:ten}
\end{equation}
showing that the velocity of electron is a periodic function of the momentum. The electron surface current density $j_z$  along the CNs axis is also given by the expression  
\[
	j_z = \frac{2e}{(2\pi \hbar)^2}\int\int f(p,t)  v_z (p) d^2 p
\]
or
\begin{equation}
	j_z = \frac{2e}{(2\pi \hbar)^2}\sum_{s=1}^{n}\int_0^{\frac{2\pi}{a}} f\big(p_z, s\Delta p_{\varphi}, \Phi(t) \big)  dp_z \label{eq:eleven}
\end{equation}                                 
where the integration is carried over the first Brillouin zone. Substituting Eqns.\eqref{eq:six}, \eqref{eq:eight} and \eqref{eq:ten} into \eqref{eq:eleven} we find the current density for the CNs after averaging over a period of time $t$, as
\begin{equation}
	j_z = \frac{8\pi e}{\sqrt{3}\hbar na_{c-c}}\sum_{r = 1}^{\infty}r\sum_{s=1}^{n}f_{rs}\epsilon_{rs}\sum_{k=-\infty}^{\infty}\frac{J_k^2(r\beta)(\Omega r + k\omega)\tau}{1 + (\Omega r + k\omega)^2\tau^2}. \label{eq:twelve}
\end{equation}
Equation \eqref{eq:twelve} can be expressed in the form
\begin{eqnarray}
	j_z &=& \frac{8\pi e}{\sqrt{3}\hbar na_{c-c}}\sum_{r = 1}^{\infty}r\Big[\frac{J_k^2(r\beta)}{1 + (\Omega r\tau)^2} + \ldots\nonumber\\
			&&2\Omega r\tau\sum_{k=1}^{\infty}\frac{J_k^2(r\beta)[1 + (\Omega r\tau)^2 - (k\omega\tau)^2]}{[1 + (\Omega r - k\omega)^2\tau^2][1 + (\Omega r + k\omega)^2\tau^2]}\Big] \nonumber\\
			&&\times\sum_{s=1}^{n}f_{rs}\epsilon_{rs}. \label{eq:thirteen}
\end{eqnarray}
If $k\omega\tau >> \Omega\tau$ and $k\omega\tau >> 1$, Eqn. \eqref{eq:thirteen} takes the form
\begin{eqnarray}
	j_z &=& \frac{8\pi e}{\sqrt{3}\hbar na_{c-c}}\sum_{r = 1}^{\infty}r\Big[\frac{J_k^2(r\beta)}{1 + (\Omega r\tau)^2} - \frac{2\Omega r\tau}{(\omega\tau)^2}\sum_{k=1}^{\infty}\frac{J_k^2(r\beta)}{k^2}\Big]\nonumber\\
	&& \times\sum_{s=1}^{n}f_{rs}\epsilon_{rs}. \label{eq:fourteen}
\end{eqnarray}	
Where  $r\beta = earE_1/\omega$, $J_0 (r\beta)$ is the Bessel function of the zeroth order, $\omega = eaE_0$ for zigzag CNs and $\Omega = eaE_0/\sqrt{3}$ for armchair CNs.

The second term in the bracket of equation \eqref{eq:fourteen} is less than the first term; however when again $r\beta$ coincides with the roots of the zeroth order Bessel function, the first term disappears and the current becomes negative, i.e. it flows against the applied d.c field. This phenomenon is called absolute negative conductivity and was first observed by Kryuchkov {\it et al.} \cite{6}.

Now, substituting $E_0 = 0$ in Eqn. \eqref{eq:eight}, we obtain
\begin{equation}
	\Phi_{\nu}(t) = \sum_{k=-\infty}^{\infty}\sum_{\nu=-\infty}^{\infty}\frac{J_k(r\beta)J_{k-\nu}(r\beta)}{1 - i(\nu\omega)\tau}exp(i\nu\omega t). \label{eq:fifteen}
\end{equation}
Hence, we obtain
\begin{equation}
j_z= \frac{8e\gamma_0}{\sqrt{3}\hbar na_{c-c}} \sum_{\nu=-\infty}^{\infty}rJ_0 (r\beta)sin(r\beta sin\omega t)\sum_{s=1}^nf_{rs} \epsilon_{rs}. \label{eq:sixteen}
\end{equation}
Where $r\beta = earE_1/\omega$, $J_{\nu}$ is the Bessel function of the $\nu^{th}$ order. For $\nu = 0$, Eq. \eqref{eq:sixteen} becomes
\begin{equation}
j_z= \frac{8e\gamma_0}{\sqrt{3}\hbar na_{c-c}} \sum_{r=1}^{\infty}rJ_0 (r\beta)sin(r\beta sin\omega t)\sum_{s=1}^nf_{rs} \epsilon_{rs}. \label{eq:seventeen}
\end{equation}

\begin{figure}[ht!]
\includegraphics[scale=.8]{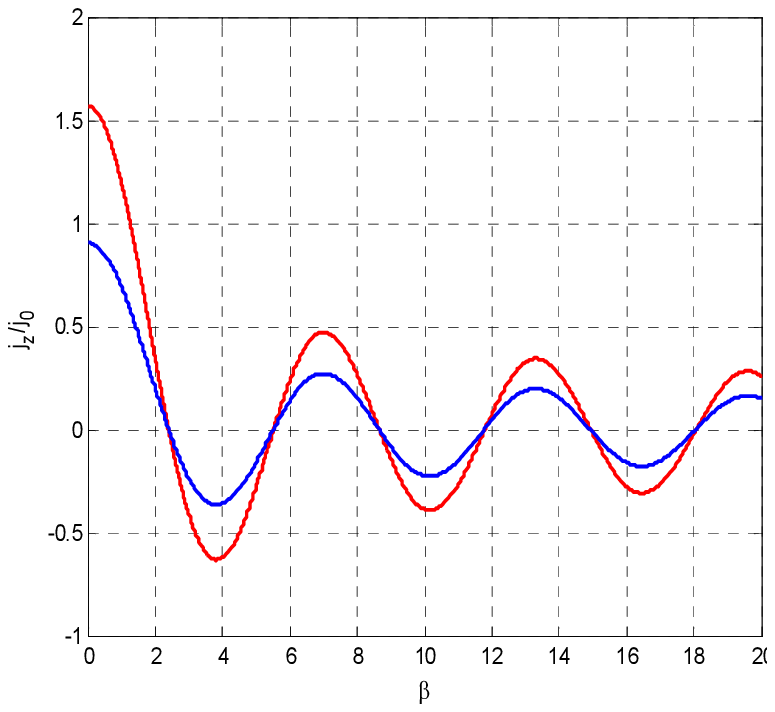}
\caption{Normalised current density ($j_z/j_0$) –amplitude ($\beta$)   characteristic curve for expression \eqref{eq:seventeen}; armchair (blue) and zigzag (red).}\label{fig:one} 
\end{figure}

\begin{figure}[ht!]
\includegraphics[scale=.8]{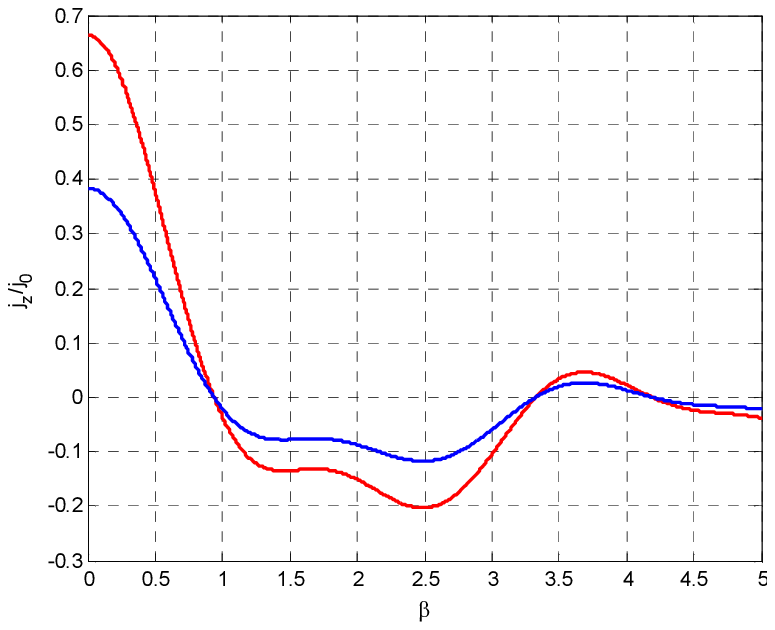}
\caption{Normalised current density ($j_z/j_0$ ) –amplitude ($\beta$)   characteristic curve for expression \eqref{eq:fifteen}; armchair (blue)  and zigzag (red) .}\label{fig:two} 
\end{figure}

\section{	Results and Discussion} 
Using the Boltzmann's transport equation with constant relaxation time, we theoretical study the electron transport phenomena in CNs. For the condition of high-frequency fields, $E(t)=E_1 cos\omega t$, an analytical expression for the current density was obtained in equation \eqref{eq:seventeen}. See Fig. \ref{fig:one}. From Eqn. \eqref{eq:seventeen} it is observed that when $r\beta$ is equal to the roots of the zeroth order Bessel function (2.4, 4.8, 8.4, 11.8, 14.8, 18.0), $j_z$ becomes zero, i.e. there is no conduction, the CNs behave as an insulator. This phenomenon is called total self induced transparency first observed by Ignatov {\it et al.} \cite{7}. However, when the CNs is exposed to an a.c. and d.c electric fields, i.e $E(t)=E_0 + E_1 cos\omega t$ and $k\omega\tau >> 1$, we obtained expression \eqref{eq:fourteen} which is an indication for absolute negative conductivity i.e. current flows against the applied d.c field. See. Fig. \ref{fig:two}.
\section{Conclusion}
In conclusion, we have theoretically studied the effect of a.c and d.c electric field on the transport properties of CNs. It is noted that in the presence of these fields, phenomena like total self induced transparency and absolute negative conductivity are observed.


\end{document}